\documentclass[pra,twocolumn,amsmath,amssymb,superscriptaddress,longbibliography]{revtex4-1}

\usepackage{bm}
\usepackage{graphicx}
\usepackage{braket}
\usepackage{url}
\usepackage{datetime}
\usepackage{stackrel}
\usepackage{booktabs}
\usepackage{array}
\usepackage{booktabs}
\usepackage{xcolor}

\newcommand{\bea}{\begin{eqnarray}}
\newcommand{\eea}{\end{eqnarray}}
\newcommand{\dt}{\partial_t}

\newcommand{\dx}{\partial_x}

\newcommand{\bwe}{\begin{widetext}\begin{eqnarray}}
\newcommand{\ewe}{\end{eqnarray}\end{widetext}}
\newcommand{\bi}{\begin{itemize}}
\newcommand{\ei}{\end{itemize}}
\newcommand{\degree}{$^{\circ}\,$}

\newcommand{\Omac}{\Omega_{ac}}
\newcommand{\Tac}{\tau_{ac}}

\begin{document}

\title{Entropy mode driven gas optics}

\author{P. Michel}
\author{A. Oudin}
\affiliation{Lawrence Livermore National Laboratory, Livermore, CA 94551, USA}
\author{H. Rajesh}
\author{K. Ou}
\author{D. Chakraborty}
\author{S. Cao}
\affiliation{Stanford University, Stanford, CA 94305, USA}
\author{E. Kur}
\affiliation{Lawrence Livermore National Laboratory, Livermore, CA 94551, USA}
\author{L. Lancia}
\affiliation{LULI -- CNRS, CEA, Sorbonne Universit\'e, Ecole Polytechnique, Institut Polytechnique de Paris, F-91128 Palaiseau, France}
\author{D. Ghosh}
\affiliation{Lawrence Livermore National Laboratory, Livermore, CA 94551, USA}
\author{C. Riconda}
\affiliation{LULI, Sorbonne Universit\'e, CNRS, Ecole Polytechnique, CEA, F-75252 Paris, France}
\author{J. S. Wurtele}
\affiliation{Department of Physics, University of California at Berkeley, Berkeley, California 94720-7300, USA}
\author{M. R. Edwards}
\affiliation{Stanford University, Stanford, CA 94305, USA}

\begin{abstract}
We propose a novel class of gaseous diffractive optical elements created by imprinting an entropy mode in a gas. Previous approaches to gaseous diffractive optics relied on the simultaneous excitation of a standing acoustic wave and an entropy mode to produce one-dimensional periodic structures. However, the presence of acoustic oscillations in the gas imposes stringent constraints on some operational parameters of these optical elements, such as their lifetime and diffraction angle. In this work, we introduce a new approach that eliminates the acoustic mode, relying solely on the entropy mode. This enables control of the lifetime and temporal profile of gaseous optical elements, and also allows the creation of arbitrary structures with greater contrast, including non-periodic patterns such as chirped gratings or lenses. This approach should allow operation over a wider parameter space, including larger diffraction angles and compatibility with laser pulse durations ranging from femtoseconds to microseconds.
\end{abstract}

\maketitle

\section{Introduction}

Laser-induced transient gratings have long been used as a diagnostic tool. The concept relies on imprinting an index modulation in a medium, such as a gas, via the absorption of a spatially-modulated laser, and then probing the resulting grating using another laser beam incident at the Bragg angle. The temporal oscillations and frequency shift of the diffracted signal provide a direct, {\textit{in situ}} measurement of the gas temperature and flow velocity \cite{EichlerBook,StampanoniAPB05,StampanoniAPB05b,CummingsAO95,EichlerJAP73}. 

It was recently demonstrated that such gas gratings can be optimized and reach high diffraction efficiencies ($>$90\%), comparable to traditional solid-state optics such as dielectric mirrors \cite{MichineCP20,MichelPRA24,MichinePF24,OuSPIE25,MatteoCLEO25}. The very high optical damage threshold of gases (above 1 kJ/cm$^2$, several orders of magnitude beyond the damage threshold of solid optical elements) and their transient nature make these novel optical elements attractive for a wide range of applications, including laser machining, welding, lithography, and as final optics for future inertial fusion energy (IFE) facilities where the survivability of solid optics poses a significant challenge \cite{GarozNF13,MoirFED00}.

Two methods have now been demonstrated for the generation of high diffraction efficiency gas gratings: i) via the absorption of modulated ultraviolet (UV) light in a gas containing a small fraction of ozone, which launches high amplitude waves with an index modulation proportional to the density modulation from the wave \cite{MichineCP20,MichelPRA24,MichinePF24,OuSPIE25}; and more recently, ii) via the absorption of $\approx$ 4 $\mu$m light in CO$_2$ gas and subsequent excitation of vibrationally-excited states of CO$_2$, relying on the variation of the refractive index with the vibrational energy of the CO$_2$ molecules \cite{MatteoCLEO25}. For the former method, which is the context of this paper, the absorption of the spatially modulated UV light (with wavelength $\lambda_\textsc{uv} \in$ [200–300] nm and modulation period $\Lambda$ typically of tens of $\mu$m) leads to localized gas heating. The acoustic period $\Tac = \Lambda / C_s$, where $C_s$ is the sound speed, is typically tens to a hundred ns at room temperature, which is much longer than the duration of the UV imprint beam (few ns). This results in the impulse-like excitation of a standing acoustic wave, which repeats itself in time at the acoustic period, as well as a stationary entropy mode, which forms a static imprint of the UV beam's intensity modulation pattern at $\Lambda$. The acoustic and entropy modes each contribute half of the maximum amplitude of the density modulation \cite{MichelPRA24}. 

Although this method demonstrated remarkable results in the laboratory, including gratings with diffraction efficiencies nearing 100\% and damage thresholds exceeding 1 kJ/cm$^2$, it also imposes significant constraints on the design and applicability of these optical elements. First, because the density modulation oscillates in time between zero and its maximum amplitude, the grating's lifetime is directly tied to the acoustic period: diffraction is only possible when the density modulation reached its peak, which only lasts for a fraction of $\Tac$. Consequently, achieving a sufficiently long grating lifetime to manipulate arbitrary pulse durations (e.g., nanosecond pulses for IFE applications) requires a large grating period $\Lambda = \Tac C_s$. This, in turn, limits the grating's application to very small diffraction angles $\psi_d$, as dictated by the Bragg condition $\psi_d=2 \text{arcsin}(\lambda_d/2\Lambda)$ where $\lambda_d$ is the wavelength of the diffracted light. The temporal oscillations of the grating may also be undesirable for some potential applications like pulse gating. Second, for a gas with uniform temperature (and thus sound speed), the standing acoustic wave can only reproduce the initial intensity pattern of the imprint beam if the pattern is purely one dimensional (1D) periodic, i.e., a 1D diffraction grating. For non-periodic structures, the acoustic period $\Tac$ varies spatially, preventing the recreation of the initial intensity pattern with 100\% contrast. This limitation could reduce the performance of more complex diffractive structures, such as diffractive lenses.

Here, we propose a new concept for gas optics that relies solely on the excitation of an entropy mode, without the involvement of acoustic modes. Using theory, simulations, and experiments, we demonstrate that if the duration of the imprint UV beam is long compared to a characteristic acoustic period, it is possible to excite an entropy mode with no  acoustic oscillations. This technique provides much greater control of the temporal profile and lifetime of the optical elements, without residual temporal oscillations. In particular, we have identified a regime where the temporal profile (and thus the lifetime) of the optical elements can mimic the UV imprint beam's, allowing arbitrary shaping of the optics' temporal profile. Furthermore, since the entropy mode does not propagate, this method allows imprinting arbitrary structures in the gas, either through the interference of UV light waves or by applying an amplitude mask. This technique should enable the creation of arbitrary volume holographic elements, such as chirped gratings, lenses and beam shapers, with higher contrast and efficiency compared to when acoustic waves are present.

The remainder of the paper is organized as follows: Sec. \ref{sec:theory} presents the linear fluid theory of the acoustic vs. entropy modes, and derives criteria for eliminating the acoustic mode as well as the two different regimes (weakly vs. strongly damped) for the entropy mode. Section \ref{sec:params} summarizes the different regimes of operation of gas gratings. Section \ref{sec:expts} presents experiments that demonstrate that the entropy mode can be isolated without acoustic modes, and that the lifetime of the resulting grating can be controlled via the imprint beam duration. Integrated simulations are presented in Sec. \ref{sec:sims}; these show excellent agreement with the experiments and demonstrate the benefits of entropy modes to excite non-periodic structures, such as chirped gratings. Section \ref{sec:conclusion} concludes the paper.

\section{Theory\label{sec:theory}}

\subsection{Linearized fluid equations}

Our theoretical model of entropy mode driven gas optics is based on the Navier-Stokes fluid equations, whose linearized expression in 1D is:
\bea
\dt \rho_1 + \rho_0 \dx u_1 &=& 0 \,, \label{eq:NS1} \\
\rho_0 \dt u_1 &=& -\dx p_1 + \frac 43 \mu \dx^2 u_1  \, ,   \label{eq:NS2} \\
\rho_0 T_0 \dt s_1 &=& \kappa \dx^2 T_1 + Q \, , \label{eq:NS3}
\eea
where $\rho$, $u$, $p$, $T$ and $s$ are the mass density, velocity, pressure, temperature and entropy, respectively. 
Quantities have been linearized to first order, e.g., $\rho=\rho_0+\rho_1$ with $|\rho_1|\ll \rho_0$. The term $Q(x,t)$ represents the heat source provided by the UV beam absorption, and $\mu$, $\kappa$ are the dynamic viscosity and thermal conductivity. Note that the contribution of ozone to the gas dynamics is neglected, since it is only present in very small amounts in typical experiments ($\sim$ [1--5]\%) to absorb the UV imprint beam energy and convert it to heat ($Q$) \cite{MichelPRA24}. The fluid described here represents the buffer gas used in experiments, typically oxygen. Some of the experiments presented later involve gas mixtures (O$_2$ / CO$_2$ mixture, cf. Sec. \ref{sec:expts}), but for simplicity we present the theory for a single species. The refractive index $n$ is directly related to the density via $n-1\propto \rho$; defining $n_0$ as the background index such that $n_0-1\propto \rho_0$ and $\delta n=n-n_0$ leads to $\rho_1/\rho_0=\delta n/(n_0-1)$. 

Combining these fluid equations with the expression for the entropy,
\bea
s_1=c_v[T_1/T_0+(1-\gamma)\rho_1/\rho_0] \, ,  \label{eq:s1}
\eea
and the ideal gas law $p=R_s\rho T$, which in the linearized form is simply
\bea
\frac{p_1}{p_0}=\frac{\rho_1}{\rho_0}+\frac{T_1}{T_0} \, ,  \label{eq:idealgas}
\eea 
provides a set of five equations for the five unknowns $\rho_1$, $u_1$, $p_1$, $s_1$ and $T_1$ ($\gamma=c_p/c_v$ is the adiabatic index, with $c_p$ and $c_v$ the specific heat capacities at constant pressure and volume, and $R_s=c_p-c_v$ is the specific gas constant).

Eliminating variables to keep only $p_1$ and $s_1$ in the system of equations above and solving the system in Fourier space leads to a system of two independent equations (wave and diffusion equations) for the pressure and the entropy \cite{PierceBook}:
\bea
(\dt^2+2\nu_{ac}\dt - C_s^2\dx^2)p_1 &=& (\gamma-1)\dt Q \,, \label{eq:acousticmode} \\
\left ( \dt - \frac{\kappa}{c_p\rho_0}\dx^2 \right)s_1 &=& \frac{Q}{\rho_0T_0} \,, \label{eq:entropymode}
\eea
where $C_s$ is the sound speed, given by $C_s^2=\gamma p_0/\rho_0$, and the acoustic wave damping $\nu_{ac}$ for a spatial mode proportional to $e^{ikx}$ is
\bea
\nu_{ac} = k^2/(2\rho_0)[4\mu /3 + (\gamma-1)\kappa/c_p] \,.
\eea

Equations \eqref{eq:acousticmode} and \eqref{eq:entropymode} represent the acoustic mode (pressure perturbation at constant entropy) and entropy mode (entropy perturbation at constant pressure), which constitute the two fundamental solutions of the linearized fluid equations in 1D (in higher dimensions a third fundamental mode is also present, the vorticity mode \cite{PierceBook}). The three remaining hydrodynamics quantities $\rho_1$, $u_1$ and $T_1$ can be derived from $p_1$ and $s_1$ via Eqs. \eqref{eq:NS1}, \eqref{eq:s1} and \eqref{eq:idealgas}. In particular, the total density perturbation resulting from the heat source is obtained by combining Eqs. \eqref{eq:s1} and \eqref{eq:idealgas}, leading to
\bea
\frac{\rho_1}{\rho_0} = \frac{1}{\gamma} \frac{p_1}{p_0}-\frac{s_1}{c_p} \,,  \label{eq:rho1acent}
\eea
where the first term on the right-hand side is associated with the acoustic mode, and the second, with the entropy mode.

The concept of entropy mode gas optics relies on heating the gas slowly compared to the acoustic time scale, thereby suppressing the acoustic transient and isolating the entropy mode. In other words, we want to make the right-hand side of Eq. \eqref{eq:acousticmode} negligible, while still exciting an entropy mode via the right-hand side of Eq. \eqref{eq:entropymode}. The next subsections present analytical solutions to the wave and diffusion equations in order to quantify this concept and describe the different regimes of the entropy mode. The solutions are derived for a heat source term with the following simplified form:
\bea
Q(x,t) = \frac{q_0}{\sqrt{2\pi}} \left[1+\cos(K x)\right] e^{-t^2/2\tau^2} \, , \label{eq:Q}
\eea
with $K=2\pi/\Lambda$. This is a reasonable approximation of the heating from a UV imprint beam with a Gaussian temporal profile of duration $\tau$, spatially modulated in space with the period $\Lambda$. The separability of time and space is not rigorously satisfied in experiments due to the nonlinear depletion of ozone and heating from secondary chemical reactions over time scales longer than the imprint beam duration \cite{MichelPRA24}. However, the approximation is sufficient to derive the main regimes of interest, as will be confirmed by experiments and integrated simulations that include these nonlinear effects.

\subsection{Solution for the acoustic mode}

For this discussion we will ignore the acoustic damping, which will be irrelevant for the regime of interest. The Green's function for the wave equation Eq. \eqref{eq:acousticmode} in the absence of damping is $G(x,t) = H(t-|x|/C_s)/2C_s$, where $H$ is the Heaviside step function. Convoluting this function with the heat source from Eq. \eqref{eq:Q} gives the following result for the pressure perturbation:
\bea
\frac{p_1(x,t)}{p_0} = \frac{\Delta T}{T_0} \Big \{ \frac 12 \left [\text{erf}(t/\sqrt{2}\tau)+1 \right] +\cos(K x) \nonumber \\
  \times \int_{-\infty}^\infty \frac{h(u)}{\tau} H(t-u) \cos\left[ \Omac(t-u)\right] du \Big \}  \label{eq:p1full}
\eea
where erf is the error function, $\Omac = 2\pi/\Tac$, $h(t)=e^{-t^2/2\tau^2}/\sqrt{2\pi}$, $\Delta T=q_0\tau /(\rho_0 c_v)$ is the total time- and space-averaged temperature increase in the gas without dissipation, and $T_0$ is the initial background temperature. The first term in square brackets in Eq. \eqref{eq:p1full} corresponds to the spatially-averaged pressure increase due to the rise in the gas temperature; the second term in the curly brackets ($\propto \cos(K x)$) is the standing wave.

	\subsubsection{Fast heating ($\Omac\tau\ll 1$)}

When the heating time $\tau$ is very small compared to the acoustic period, such that $h(u)\rightarrow \tau \delta(u)$ and $[\text{erf}(t/\sqrt{2}\tau)+1]/2\rightarrow H(t)$, the expression for the pressure perturbation becomes
\bea
\left. \frac{p_1(x,t)}{p_0} \right|_{\Omac\tau\ll 1} \approx H(t)\frac{\Delta T}{T_0}\left[1+\cos(Kx)\cos(\Omac t) \right] . \label{eq:p1fast}
\eea

The rapid heating launches a standing acoustic wave, as was already discussed in Ref. \cite{MichelPRA24} and observed in previous experiments \cite{MichineCP20,MichinePF24,OuSPIE25}.

	\subsubsection{Slow heating ($\Omac\tau\gg 1$)}

We now turn to the slow heating regime, to eliminate the standing acoustic wave. For times well past the end of the heating pulse, $t\gg \tau$, Eq. \eqref{eq:p1full} yields \footnote{For $t\gg \tau$, $h(u)H(t-u)\approx h(u)$, and the integral in Eq. \eqref{eq:p1full} takes the form of a convolution, $h(t)*\cos(\Omac t)/\tau$. Using the convolution theorem, $h(t)*\cos(\Omac t)/\tau=\mathcal{F}^{-1}\{ \mathcal{F}[h(t)/\tau] \mathcal{F}[\cos(\Omac t)] \}/\sqrt{2\pi}$, with $\mathcal{F}$ the Fourier transform (unitary definition, with angular frequency), and $\mathcal{F}[h(t)/\tau]=e^{-\omega^2\tau^2/2}$, $\mathcal{F}[\cos(\Omac t)]=\sqrt{\pi/2}[\delta(\omega-\Omac)+\delta(\omega+\Omac)]$, leads to Eq. \eqref{eq:p1late}.}:
\bea
\frac{p_1(x,t\gg \tau)}{p_0} &\approx& \frac{\Delta T}{T_0} \left[ 1+e^{-\Omac^2\tau^2/2}\cos(K x)\times \right . \nonumber \\
&&  \cos(\Omac t) \Big ] \,. \label{eq:p1late}
\eea

The ``1'' in the square bracket represents the final spatially-averaged pressure increase from gas heating, and the rest is the standing wave, whose amplitude scales like $\exp[-\Omac^2\tau^2/2]$. Therefore, the contribution from the standing acoustic wave can be made negligible if $\tau \gg 1/\Omac$ (the ``slow heating'' criterion mentioned earlier), which is the intended outcome. This can be achieved by increasing the duration of the imprint pulse and/or reducing the acoustic period, as will be discussed in Secs. \ref{sec:params} and \ref{sec:expts}. In this limit, the oscillations in the integral in Eq. \eqref{eq:p1full} make that integral become infinitely small (as can be shown by integrating by parts), and the pressure perturbation simply becomes
\bea
\left . \frac{p_1}{p_0} \right |_{\Omac\tau \gg 1} \approx \frac{\Delta T}{2T_0} \left [\text{erf}(t/\sqrt{2}\tau)+1 \right] \,, \label{eq:p1slow}
\eea
i.e., a spatially uniform pressure increase due to the gas heating without acoustic oscillations.

\subsection{Solution for the entropy mode}

Eliminating the standing acoustic wave by setting $\tau \gg 1/\Omac$ can be achieved while still obtaining a strong density modulation from the entropy mode in the gas. The solution for the entropy mode can be obtained via the Green's function for the diffusion equation, Eq. \eqref{eq:entropymode}. The Green's function $G(x,t)=H(t)e^{-x^2/4Dt}/\sqrt{4\pi Dt}$, where $D=\kappa/c_p\rho_0$ is the diffusion coefficient, can be convolved with the heat source from Eq. \eqref{eq:Q} to give the solution:
\bea
s_1 =& \dfrac{c_v}{2}\dfrac{\Delta T}{T_0} \Big \{ \text{erf}(t/\sqrt{2}\tau)+1+e^{\nu_e^2\tau^2/2-\nu_e t}\cos(K x) \nonumber \\
   &\times \left[ \text{erf}\left (t/\sqrt{2}\tau - \nu_e\tau/\sqrt{2} \right) +1 \right] \Big \} \,, \label{eq:s1full}
\eea
with $\nu_e=K^2 D$ the entropy mode damping rate. Two different regimes can be identified for the entropy mode, depending on the ratio of the damping time to UV imprint beam duration, as described below.

	\subsubsection{Weakly damped entropy mode ($\nu_e \tau \ll 1$)}

The expression for the density modulation of the weakly damped entropy mode for slow heating ($\Omac\tau \gg 1$) is obtained by taking the limit of $\nu_e \tau \ll 1$ in Eq. \eqref{eq:s1full}, as well as $p_1/p_0$ from Eq. \eqref{eq:p1slow}, and inserting the resulting expressions for $p_1$ and $s_1$ in Eq. \eqref{eq:rho1acent}, leading to
\bea
\left . \frac{\rho_1}{\rho_0} \right|_{\nu_e \tau \ll 1} \approx \frac{-\Delta T}{2\gamma T_0} \left [\text{erf}(t/\sqrt{2}\tau)+1 \right] e^{-\nu_et}\cos(K x)  \,.  \label{eq:rho1entwd}
\eea

In this regime, the grating forms over a time $\sim \tau$ (term in the square bracket) and stays on for a time $\sim 1/\nu_e \gg \tau$ (exponential term). In other words, the grating lifetime is dictated by thermal diffusion and is on the order of the damping time $1/\nu_e$. Isolating the entropy mode in this regime (without the acoustic mode) thus requires $1/\Omac \ll \tau \ll 1/\nu_e$.

	\subsubsection{Strongly damped entropy mode ($\nu_e \tau \gg 1$)}

Taking the opposite limit of $\nu_e \tau \gg 1$ in Eq. \eqref{eq:s1full} and combining with Eqs. \eqref{eq:p1slow} and \eqref{eq:rho1acent} gives the expression for the density perturbation from the strongly damped regime of the entropy mode \footnote{Here one must use the lowest order of the asymptotic expansion of the error function, $\text{erf}(x) \approx 1-\exp[-x^2]/(\sqrt{\pi}x)$ when $x\rightarrow \infty$.}:
\bea
\left . \frac{\rho_1}{\rho_0} \right |_{\nu_e \tau \gg 1} = \frac{-\Delta T}{2\gamma T_0} \sqrt{\frac{2}{\pi}}\frac{1}{\nu_e \tau} \cos(K x)e^{-t^2/2\tau^2} \,. \label{eq:rho1entsd}
\eea

Now, the temporal behavior of the grating---and thus its lifetime---is exactly the same as the imprint pulse's ($\propto \exp[-t^2/2\tau^2]$). The entropy mode's damping due to diffusion occurs faster than the duration of the UV imprint beam; the mode can still exist in a ``forced'' state but quickly vanishes once the imprint beam is turned off.

The maximum density modulation between the weakly vs. strongly damped regimes differs by a factor proportional to $\nu_e \tau$. This means that achieving the same diffraction efficiency in the strongly damped regime as in the weakly damped regime requires  increasing the gas heating $\Delta T$ (by adjusting the ozone fraction or the imprint beam fluence) and/or increasing the grating thickness (as the diffraction efficiency depends on the factor $\delta n L_g/\lambda_d$, with $L_g$ the grating thickness and $\delta n=(n_0-1)\rho_1/\rho_0$ the index modulation, with $n_0$ the average background index).


\section{Parameter space \label{sec:params}}

\begin{figure}[htbp]
\includegraphics{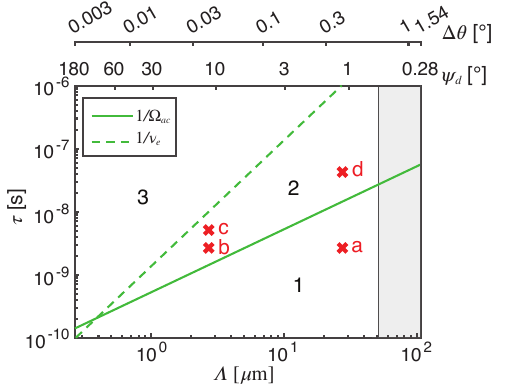}
\caption{Parameter space for gas gratings, in terms of the UV imprint pulse duration $\tau$ and modulation period $\Lambda$. The regions marked 1, 2 and 3 correspond to the regimes of acoustic and entropy modes, weakly damped isolated entropy mode, and strongly damped isolated entropy mode, respectively (cf. table \ref{tab:param}). The markers represent the parameters for the experiments and simulations shown in Figs. \ref{fig:expts} and \ref{fig:PiafsEntropyWD}, respectively; the grey area corresponds to the regime of diffraction into higher-order modes for $\rho_1/\rho_0=0.4$. $\Delta \theta$ and $\psi_d$ (upper axes) are the grating's angular bandwidth (also for $\rho_1/\rho_0=0.4$) and the diffraction angle, respectively, assuming 532 nm diffracted light.}
\label{fig:param}
\end{figure}

The parameter space of gas gratings, summarizing the key results from the theoretical discussion above, is represented in Fig. \ref{fig:param}. The figure shows $1/\Omac$ and $1/\nu_e$ (acoustic period and entropy mode damping time) as a function of the grating wavelength $\Lambda$ for a sound speed $C_s$ = 302 m/s and a thermal diffusivity of 2$\times 10^{-5}$ m$^2$/s (the value for oxygen at room temperature).

Three different regimes can be identified, labelled 1--3 in the figure:
\begin{enumerate}
\item $\tau \ll 1/\Omac, 1/\nu_e$: acoustic and entropy modes (fast heating), weakly damped. The grating density modulation is obtained from Eqs. \eqref{eq:p1fast}, Eq. \eqref{eq:s1full} with $\nu_e \tau \ll 1$ and $\tau \rightarrow 0$, and Eq. \eqref{eq:rho1acent}, leading to
\bea
\frac{\rho_1}{\rho_0} \approx H(t)\frac{\Delta T}{\gamma T_0}\left[1+\cos(Kx)\left(\cos(\Omac t)-1\right) \right] \,,
\eea
as was previously derived in Ref. \cite{MichelPRA24}.

\item $1/\Omac \ll \tau \ll 1/\nu_e$: isolated entropy mode, weakly damped. The density perturbation is given by Eq. \eqref{eq:rho1entwd}.

\item $\tau \gg 1/\Omac, 1/\nu_e$: isolated entropy mode, strongly damped: the density modulation is given by Eq. \eqref{eq:rho1entsd}.
\end{enumerate}

\begin{table*}[t]\centering
  \renewcommand{\arraystretch}{1.5}
    \begin{tabular}{|l | l | l  | l |}
    \hline
     Region & 1 (acoustic \& entropy modes) & 2 (entropy mode, weakly damped) & 3  (entropy mode, strongly damped) \\
    \hline
    Regime & $\tau \ll \Tac, 1/\nu_e$ & $\Tac \ll \tau \ll \frac{1}{\nu_e}$ & $\tau \gg \Tac, 1/\nu_e$ \\
    Grating lifetime \, & $\approx \Tac/4$ & $\approx 1/\nu_e$ & $\approx \tau$ \\
    Max. $\rho_1/\rho_0$ & $(2/\gamma)\Delta T/T_0$  & $(1/\gamma)\Delta T/T_0$  & $(\sqrt{2\pi}\nu_e\tau \gamma)^{-1}\Delta T/T_0$ \\
     \hline
    \end{tabular}
    \caption{Summary of the different regimes of gas optics, with associated grating lifetimes and maximum density modulation amplitudes.}
    \label{tab:param}
\end{table*}

The upper axes show the diffraction angle (angle between the incident and diffracted beams) $\psi_d=2$arcsin$(\lambda_d/2\Lambda)$, and the grating's angular bandwidth $\Delta \theta=(n_0-1)(\rho_1/\rho_0) \Lambda/\lambda_d$ (the angular bandwidth is the maximum angular deviation from the Bragg angle before the diffraction efficiency starts to drop \cite{YarivBook}), shown here for $\rho_1/\rho_0=0.4$. Both $\psi_d$ and $\Delta \theta$ were calculated for a diffracted wavelength of 532 nm. The bandwidth increases as the diffraction angle decreases, which can be controlled via $\Lambda$. At some point, the bandwidth becomes as large as the diffraction angle: this is where the transition from the Bragg (diffraction into the first-order mode only) to the Raman-Nath (diffraction into higher-order modes) regimes of diffraction occurs. The criteria that best captures that transition is the parameter $\alpha = \lambda_d^2/(\Lambda^2 n_0\delta n)$, with $\delta n=(n_0-1)\rho_1/\rho_0$ \cite{MoharamAO78}: $\alpha \gg 1$ corresponds to the Bragg diffraction regime, and $\alpha \ll 1$ to the Raman-Nath regime. The Raman-Nath regime, presumably unwanted if one aims to use the grating as a dielectric mirror, is represented by the shaded grey region on the figure for $\rho_1/\rho_0=0.4$. At the other limit of small grating period $\Lambda$, the plot stops at $\Lambda=\lambda_d/2$ (where $\lambda_d$ = 532 nm is the diffracted light wavelength), which represents the minimum grating period for which $\psi_d$ = 180\degree (reflection at normal incidence).

Another important parameter for applications is the lifetime of the grating, which depends on the gas optics regime, as summarized in table \ref{tab:param}. In region 1, the grating lifetime is dictated by the acoustic oscillations and is approximately $\Tac/4$, as described in Ref. \cite{MichelPRA24}. Whereas in regions 2 and 3, the lifetime is dictated by the entropy mode damping time $1/\nu_e$ and the imprint beam duration $\tau$, respectively. Designing gas optics with only the entropy mode allows greater control over their lifetime and temporal behavior, without residual acoustic oscillations.

We also note that another option for reducing acoustic oscillations and isolating the entropy mode involves driving very high-amplitude, nonlinear acoustic waves with a shock formation time shorter than the acoustic damping time. In this regime, the nonlinear damping of acoustic oscillations due to wave steepening does not affect the entropy mode, since the entropy mode does not propagate. As a result, the acoustic oscillations are rapidly damped, while the density modulation from the entropy mode remains, as discussed in Ref. \cite{OudinPOP25} and demonstrated in Ref. \cite{OuSPIE25}. Although both approaches have now been shown to be effective, the method presented here (operating in the $\Omac\tau\gg 1$ regime) should offer greater flexibility in terms of experimental parameters.

The table also lists the expressions of the maximum density modulation for the three regimes. As mentioned earlier, entropy mode optics, particularly in the strongly damped regime, are expected to require more energy in the UV imprint beam (i.e., a higher temperature increase $\Delta T$) or a thicker volume of gas (along the direction of the diffracted laser beam) to achieve the same diffraction efficiency.

Designing gas optics in the entropy mode regime can also open the design space of these elements to larger diffraction angles and long pulse durations. On the other hand, it should be kept in mind that increasing the diffraction angle reduces the the grating bandwidth, which imposes a stricter angular alignment and may reduce the diffraction efficiency for broadband lasers, depending on the exact parameters (the grating's spectral bandwidth is $\Delta \omega/\omega_d=2(\delta n/n_0)(\Lambda/\lambda_d)^2$).

\section{Experiments\label{sec:expts}}

Experiments were performed at Stanford University using a 1 J Nd:YAG (Spectra-Physics PIV-400) Q-switched laser system. The laser was frequency-quadrupled from 1064 nm to 267 nm to generate a UV beam. The UV beam was split into a pair of equal-intensity imprint beams made to overlap with an adjustable crossing angle at the exit of a gas flow tube. A resin tube with a rectangular cross section directed a flow of O$_3$, generated with a corona discharge ozone generator. The O$_3$ was buffered with a mixture of O$_2$ and CO$_2$ such that the fraction of O$_3$ was a few percent. This flow had a height of 10 mm, a thickness of 2.5 mm, and a flow rate ranging from 1.5 to 3.5 LPM. Although CO$_2$ can enhance diffraction efficiency by increasing gas heating, its primary role in these experiments was to adjust the temporal characteristics of the grating. Increasing the CO$_2$ fraction reduces the speed of sound in the mixture, increasing the gas transit time across a grating period. The flow tube design also allows for a co-flow of N$_2$ surrounding the gas mixture to limit shear-driven mixing with air. Both imprint beams were focused to a 3 mm diameter at the flow tube using an f/40 lens, yielding fluences ranging from one to several hundred mJ/cm$^2$.

\begin{figure*}[htbp]
\includegraphics{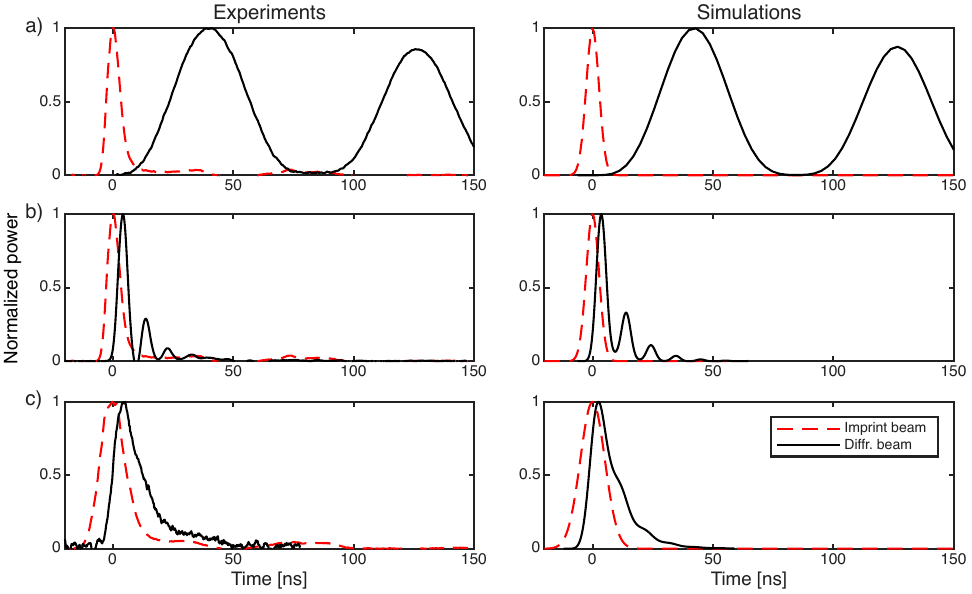}
\caption{Experimental measurements (left) and PIAFS simulations (right) showing the power (in arbitrary units) of the imprint beam (dashed) and diffracted beam (solid), illustrating the transition from the acoustic/entropy modes (a) to the isolated entropy mode (c) regime. The plots a, b and c correspond to the markers a, b and c in Fig. \ref{fig:param}, with $\Lambda$ = 27.4, 2.7 and 2.7 $\mu$m, respectively, and $\tau$ = 2.65, 2.65 and 5.1 ns, respectively (FWHM = 6.25, 6.25 and 12.0 ns; simulations assume perfect Gaussian pulse shapes). The parameter $\Omac\tau$ is equal to 0.2, 1.7 and 3.3 for a, b and c---i.e., an attenuation factor $\exp[-(\Omac \tau)^2/2]$ = 0.98, 0.24 and 0.0043 for the acoustic oscillations, per Eq. \eqref{eq:p1late}.}
\label{fig:expts}
\end{figure*}

In order to isolate the entropy mode, the grating period was reduced (to get a smaller acoustic period), and the pulse duration of the UV imprint beam was increased from 6.25 ns Gaussian full width at half maximum (FWHM) to 12.0 ns by detuning the laser Q-switch delay. Reducing the Q-switch delay lowers the laser output energy, which leads to a nonlinear drop in conversion efficiency to 266 nm. As a result, the UV fluence was too low in these experiments to achieve high diffraction efficiency due to limited heating of the gas mixture. However, the goal was to investigate the temporal behavior of the diffracted light and demonstrate the transition from the combined acoustic/entropy mode regime ($\Omac\tau \ll 1$) to the isolated entropy mode regime ($\Omac\tau \gg 1$) rather than to achieve high diffraction efficiency. 

A 532 nm continuous wave diode laser was diffracted to characterize the temporal profile of the grating.  An f/20 lens was used to focus the beam at the flow tube, thereby introducing angular spread to ensure that a portion of the beam would meet the Bragg condition and undergo diffraction despite the narrow angular bandwidth caused by a short grating period. The diffracted signal was measured with a photodetector and an oscilloscope.

Figure \ref{fig:expts} (left column) shows experimental measurements of the diffracted beam temporal profile for the parameters of the three markers a, b, and c in Fig. \ref{fig:param}, with $\Lambda$ = 27.4, 2.7 and 2.7 $\mu$m, respectively, and $\tau$ = 2.65, 2.65 and 5.1 ns, respectively (FWHM = 6.25, 6.25 and 12.0 ns). The parameter $\Omac\tau$, which measures the transition from the acoustic/entropy mode to the isolated entropy mode regime, is equal to 0.2, 1.7 and 3.3 for Figs. \ref{fig:expts}a, b and c, respectively. These values lead to an ``attenuation factor'' for the acoustic oscillations $\exp[-(\Omac \tau)^2/2]$ (cf. Eq. \eqref{eq:p1late}) equal to 0.98, 0.24 and 0.0043, respectively---i.e., ranging from the regime of full oscillations in Fig. \ref{fig:expts}a to no oscillations in \ref{fig:expts}c. While measurements in Fig. \ref{fig:expts}a used pure O$_2$ as the buffer gas, those in Figs. \ref{fig:expts}b and \ref{fig:expts}c used a buffer of 85\% CO$_2$ / 15\% O$_2$, resulting in a lower acoustic speed (302 m/s, accounting for a 18\% increase of the gas temperature due to the absorption of the imprint beam, vs. 318 m/s for Fig. \ref{fig:expts}a; this was estimated by measuring the acoustic period as it became visible again for shorter imprint beam durations). As predicted by theory, for $\Omac\tau$ = 3.3 (Fig. \ref{fig:expts}c) the UV pulse duration approaches the entropy mode damping time, resulting in a grating lifetime and temporal profile comparable to that of the imprint beam.

\section{Simulations\label{sec:sims}}

In this section, we present simulations from the code PIAFS \cite{OudinPOP25} to validate the theory and help visualize the dynamics of entropy mode gas optics. PIAFS is an integrated tool that combines the chemistry of gas heating via absorption of UV light and the nonlinear gas dynamics by solving the full Navier-Stokes system of equations. The space- and time-dependent heating source in the fluid model is calculated self-consistently with the hydrodynamics by solving coupled rate equations for the various chemical reactions at play, including the photodissociation of the ozone molecules (O$_3$) and subsequent chemical reactions between the dissociation products (O and O$_2$) and the background gas, as described in Ref. \cite{MichelPRA24}. The resulting density modulation is then used to compute the time-dependent diffraction efficiency of a laser incident on the resulting diffractive element using coupled wave theory or a paraxial propagation model. In the following, the diffracted laser beam is assumed to be incident at the Bragg angle.

Figure \ref{fig:expts} (right column) shows the simulated diffracted signal for the same parameters as the experimental results on the left column (markers a, b, c in Fig. \ref{fig:param}), revealing excellent agreement between experiments and simulations.

\begin{figure}[htbp]
\includegraphics{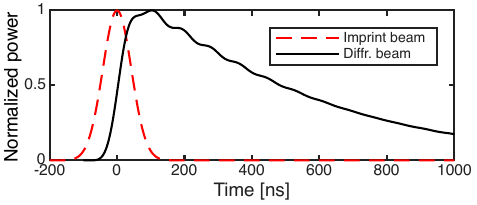}
\caption{PIAFS simulations showing the power (in arbitrary units) of the imprint and diffracted signals for the parameters represented by the marker ``d'' in Fig. \ref{fig:param} ($\Lambda$ = 27.4 $\mu$m and $\tau=42.5$ ns, longer than the acoustic period but shorter than the entropy mode damping time), illustrating the weakly damped entropy mode regime.}
\label{fig:PiafsEntropyWD}
\end{figure}

We also simulated another set of parameters in Fig. \ref{fig:PiafsEntropyWD}, corresponding to the marker ``d'' in Fig. \ref{fig:param}, with $\Lambda$ = 27.4 $\mu$m and $\tau=42.5$ ns (i.e., FWHM = 100 ns). The simulation illustrates the regime of a weakly damped entropy mode, where the lifetime of the grating is dictated by the damping time of the entropy mode---as opposed to the strongly damped regime in Fig. \ref{fig:expts}c, where the grating lifetime is approximately equal to the imprint beam's duration. This imprint pulse duration is not accessible experimentally with our current laser system, whose pulse duration is not easily adjustable. However, this regime should be easily accessible with a longer pulse laser system, and enable the formation of gas gratings with lifetimes of hundreds of nanoseconds.

To illustrate the additional advantage of using entropy modes to imprint non-periodic structures, we present simulations of a chirped grating created using both acoustic and entropy modes versus using an entropy mode alone. The grating is created via the absorption of a UV imprint beam with the following intensity profile along $x$:
\bea
I_\textsc{uv}(x,t)=I_0 e^{-t^2/2\tau^2} \left(1+\cos\left[K x(1-0.05x)\right] \right) \,,
\eea
with $\Lambda=2\pi/K=$ 30 $\mu$m. We only consider the spatial region $x\in [-2\Lambda,2\Lambda]$.

\begin{figure}[t]
\includegraphics[width=\linewidth]{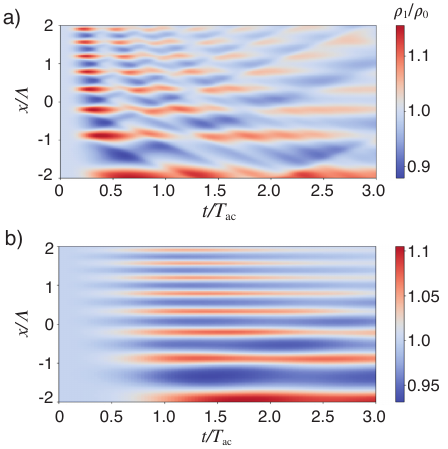}
\caption{PIAFS simulations of a chirped grating, showing the gas density modulation as a function of space and time. The imprint beam has the same chirped intensity modulation along $x$ for both cases, $I(x)\propto 1+\cos[K x(1+0.05x)]$ with $\Lambda=2\pi/K=$ 30 $\mu$m. The acoustic period used for the time axis normalization is defined with respect to the modulation period at $x=0$, i.e., $\Tac=\Lambda/C_s$. a) $\tau$ = 10 ns, $\Omac\tau=0.63$: acoustic \& entropy modes are both present, leading to reduced contrast at any given time and rapid washing out of the grating. b) $\tau$ = 100 ns, $\Omac\tau=6.3$: isolated entropy mode, showing a clean grating without acoustic transients.}
\label{fig:chirped}
\end{figure}

Figure \ref{fig:chirped}a shows a simulation for the acoustic/entropy mode regime (region 1 in Fig. \ref{fig:param}), with $\tau$ = 10 ns and $\tau\Omac=0.63$ (with $\Omac$ defined at $x=0$). 
Since the acoustic period varies along $x$, there isn't a single time at which the grating's contrast is uniform along $x$, thereby reducing the diffraction efficiency. While the imprint beam pattern is visible at early times (near $t=$ 0.25$\Tac$), it exhibits a non-uniform contrast along $x$ and becomes rapidly disrupted and washed out at later times as a broad spectrum of acoustic waves propagate through the gas.

On the other hand, Fig. \ref{fig:chirped}b illustrates the density modulation of a chirped grating generated while isolating the entropy mode, with $\tau$ = 100 ns and $\Omac\tau=6.3$. This was achieved by increasing the imprint beam duration, keeping all other parameters the same. In this case, a clean, uniform grating is formed that precisely matches the imprint beam pattern.

This approach can similarly be applied to other diffractive optics elements, particularly holographic lenses, whose index modulation pattern essentially corresponds to a chirped grating along the radial direction \cite{EdwardsPRL22}.

\section{Conclusion\label{sec:conclusion}}
We have introduced and demonstrated a new method for generating gaseous diffractive optics elements by imprinting a diffraction pattern in the gas density through the excitation of entropy modes, while suppressing transient acoustic modes. This approach enables the control of the lifetime and temporal behavior of these optical elements, as shown via theory, simulations and experiments. This scheme has the potential of enabling imprinting arbitrary structures in gases with improved contrast, facilitating the development of innovative gaseous elements such as chirped gratings or diffractive lenses.

\begin{acknowledgments}
We acknowledge stimulating discussions with Laurent Divol. This work was performed under the auspices of the U.S. Department of Energy by Lawrence Livermore National Laboratory under Contract DE-AC52-07NA27344, and was funded by the Laboratory Research and Development Program at LLNL under Project Tracking Code No. 24-ERD-001. C. R. acknowledges financial support from the Fédération de Recherche Plasmas Paris - PLAS@PAR. This work was partially supported by NNSA Grant DE-NA0004130 and NSF Grant PHY-2308641.
\end{acknowledgments}


\begin{thebibliography}{19}%
\makeatletter
\providecommand \@ifxundefined [1]{%
 \@ifx{#1\undefined}
}%
\providecommand \@ifnum [1]{%
 \ifnum #1\expandafter \@firstoftwo
 \else \expandafter \@secondoftwo
 \fi
}%
\providecommand \@ifx [1]{%
 \ifx #1\expandafter \@firstoftwo
 \else \expandafter \@secondoftwo
 \fi
}%
\providecommand \natexlab [1]{#1}%
\providecommand \enquote  [1]{``#1''}%
\providecommand \bibnamefont  [1]{#1}%
\providecommand \bibfnamefont [1]{#1}%
\providecommand \citenamefont [1]{#1}%
\providecommand \href@noop [0]{\@secondoftwo}%
\providecommand \href [0]{\begingroup \@sanitize@url \@href}%
\providecommand \@href[1]{\@@startlink{#1}\@@href}%
\providecommand \@@href[1]{\endgroup#1\@@endlink}%
\providecommand \@sanitize@url [0]{\catcode `\\12\catcode `\$12\catcode
  `\&12\catcode `\#12\catcode `\^12\catcode `\_12\catcode `\%12\relax}%
\providecommand \@@startlink[1]{}%
\providecommand \@@endlink[0]{}%
\providecommand \url  [0]{\begingroup\@sanitize@url \@url }%
\providecommand \@url [1]{\endgroup\@href {#1}{\urlprefix }}%
\providecommand \urlprefix  [0]{URL }%
\providecommand \Eprint [0]{\href }%
\providecommand \doibase [0]{http://dx.doi.org/}%
\providecommand \selectlanguage [0]{\@gobble}%
\providecommand \bibinfo  [0]{\@secondoftwo}%
\providecommand \bibfield  [0]{\@secondoftwo}%
\providecommand \translation [1]{[#1]}%
\providecommand \BibitemOpen [0]{}%
\providecommand \bibitemStop [0]{}%
\providecommand \bibitemNoStop [0]{.\EOS\space}%
\providecommand \EOS [0]{\spacefactor3000\relax}%
\providecommand \BibitemShut  [1]{\csname bibitem#1\endcsname}%
\let\auto@bib@innerbib\@empty
\bibitem [{\citenamefont {Eichler}\ \emph {et~al.}(2013)\citenamefont
  {Eichler}, \citenamefont {G{\"u}nter},\ and\ \citenamefont
  {Pohl}}]{EichlerBook}%
  \BibitemOpen
  \bibfield  {author} {\bibinfo {author} {\bibfnamefont {Hans~Joachim}\
  \bibnamefont {Eichler}}, \bibinfo {author} {\bibfnamefont {Peter}\
  \bibnamefont {G{\"u}nter}}, \ and\ \bibinfo {author} {\bibfnamefont
  {Dieter~W}\ \bibnamefont {Pohl}},\ }\href@noop {} {\emph {\bibinfo {title}
  {Laser-induced dynamic gratings}}}\ (\bibinfo  {publisher} {Springer Berlin,
  Heidelberg},\ \bibinfo {year} {2013})\BibitemShut {NoStop}%
\bibitem [{\citenamefont {Stampanoni-Panariello}\ \emph
  {et~al.}(2005{\natexlab{a}})\citenamefont {Stampanoni-Panariello},
  \citenamefont {Kozlov}, \citenamefont {Radi},\ and\ \citenamefont
  {Hemmerling}}]{StampanoniAPB05}%
  \BibitemOpen
  \bibfield  {author} {\bibinfo {author} {\bibfnamefont {A.}~\bibnamefont
  {Stampanoni-Panariello}}, \bibinfo {author} {\bibfnamefont {D.~N.}\
  \bibnamefont {Kozlov}}, \bibinfo {author} {\bibfnamefont {P.~P.}\
  \bibnamefont {Radi}}, \ and\ \bibinfo {author} {\bibfnamefont
  {B.}~\bibnamefont {Hemmerling}},\ }\bibfield  {title} {\enquote {\bibinfo
  {title} {Gas phase diagnostics by laser-induced gratings {I}. {T}heory},}\
  }\href@noop {} {\bibfield  {journal} {\bibinfo  {journal} {Applied Physics
  B}\ }\textbf {\bibinfo {volume} {81}},\ \bibinfo {pages} {101--111} (\bibinfo
  {year} {2005}{\natexlab{a}})}\BibitemShut {NoStop}%
\bibitem [{\citenamefont {Stampanoni-Panariello}\ \emph
  {et~al.}(2005{\natexlab{b}})\citenamefont {Stampanoni-Panariello},
  \citenamefont {Kozlov}, \citenamefont {Radi},\ and\ \citenamefont
  {Hemmerling}}]{StampanoniAPB05b}%
  \BibitemOpen
  \bibfield  {author} {\bibinfo {author} {\bibfnamefont {A.}~\bibnamefont
  {Stampanoni-Panariello}}, \bibinfo {author} {\bibfnamefont {D.~N.}\
  \bibnamefont {Kozlov}}, \bibinfo {author} {\bibfnamefont {P.~P.}\
  \bibnamefont {Radi}}, \ and\ \bibinfo {author} {\bibfnamefont
  {B.}~\bibnamefont {Hemmerling}},\ }\bibfield  {title} {\enquote {\bibinfo
  {title} {Gas-phase diagnostics by laser-induced gratings {II}.
  {E}xperiments},}\ }\href@noop {} {\bibfield  {journal} {\bibinfo  {journal}
  {Applied Physics B}\ }\textbf {\bibinfo {volume} {81}},\ \bibinfo {pages}
  {113--129} (\bibinfo {year} {2005}{\natexlab{b}})}\BibitemShut {NoStop}%
\bibitem [{\citenamefont {Cummings}\ \emph {et~al.}(1995)\citenamefont
  {Cummings}, \citenamefont {Leyva},\ and\ \citenamefont
  {Hornung}}]{CummingsAO95}%
  \BibitemOpen
  \bibfield  {author} {\bibinfo {author} {\bibfnamefont {E.~B.}\ \bibnamefont
  {Cummings}}, \bibinfo {author} {\bibfnamefont {I.~A.}\ \bibnamefont {Leyva}},
  \ and\ \bibinfo {author} {\bibfnamefont {H.~G.}\ \bibnamefont {Hornung}},\
  }\bibfield  {title} {\enquote {\bibinfo {title} {Laser-induced thermal
  acoustics ({LITA}) signals from finite beams},}\ }\href@noop {} {\bibfield
  {journal} {\bibinfo  {journal} {Appl. Opt.}\ }\textbf {\bibinfo {volume}
  {34}},\ \bibinfo {pages} {3290--3302} (\bibinfo {year} {1995})}\BibitemShut
  {NoStop}%
\bibitem [{\citenamefont {Eichler}\ \emph {et~al.}(1973)\citenamefont
  {Eichler}, \citenamefont {Salje},\ and\ \citenamefont
  {Stahl}}]{EichlerJAP73}%
  \BibitemOpen
  \bibfield  {author} {\bibinfo {author} {\bibfnamefont {H.}~\bibnamefont
  {Eichler}}, \bibinfo {author} {\bibfnamefont {G.}~\bibnamefont {Salje}}, \
  and\ \bibinfo {author} {\bibfnamefont {H.}~\bibnamefont {Stahl}},\ }\bibfield
   {title} {\enquote {\bibinfo {title} {{Thermal diffusion measurements using
  spatially periodic temperature distributions induced by laser light}},}\
  }\href@noop {} {\bibfield  {journal} {\bibinfo  {journal} {Journal of Applied
  Physics}\ }\textbf {\bibinfo {volume} {44}},\ \bibinfo {pages} {5383--5388}
  (\bibinfo {year} {1973})}\BibitemShut {NoStop}%
\bibitem [{\citenamefont {Michine}\ and\ \citenamefont
  {Yoneda}(2020)}]{MichineCP20}%
  \BibitemOpen
  \bibfield  {author} {\bibinfo {author} {\bibfnamefont {Yurina}\ \bibnamefont
  {Michine}}\ and\ \bibinfo {author} {\bibfnamefont {Hitoki}\ \bibnamefont
  {Yoneda}},\ }\bibfield  {title} {\enquote {\bibinfo {title} {Ultra high
  damage threshold optics for high power lasers},}\ }\href@noop {} {\bibfield
  {journal} {\bibinfo  {journal} {Communications Physics}\ }\textbf {\bibinfo
  {volume} {3}},\ \bibinfo {pages} {24} (\bibinfo {year} {2020})}\BibitemShut
  {NoStop}%
\bibitem [{\citenamefont {Michel}\ \emph {et~al.}(2024)\citenamefont {Michel},
  \citenamefont {Lancia}, \citenamefont {Oudin}, \citenamefont {Kur},
  \citenamefont {Riconda}, \citenamefont {Ou}, \citenamefont {Perez-Ramirez},
  \citenamefont {Lee},\ and\ \citenamefont {Edwards}}]{MichelPRA24}%
  \BibitemOpen
  \bibfield  {author} {\bibinfo {author} {\bibfnamefont {P.}~\bibnamefont
  {Michel}}, \bibinfo {author} {\bibfnamefont {L.}~\bibnamefont {Lancia}},
  \bibinfo {author} {\bibfnamefont {A.}~\bibnamefont {Oudin}}, \bibinfo
  {author} {\bibfnamefont {E.}~\bibnamefont {Kur}}, \bibinfo {author}
  {\bibfnamefont {C.}~\bibnamefont {Riconda}}, \bibinfo {author} {\bibfnamefont
  {K.}~\bibnamefont {Ou}}, \bibinfo {author} {\bibfnamefont {V.M.}\
  \bibnamefont {Perez-Ramirez}}, \bibinfo {author} {\bibfnamefont
  {J.}~\bibnamefont {Lee}}, \ and\ \bibinfo {author} {\bibfnamefont {M.R.}\
  \bibnamefont {Edwards}},\ }\bibfield  {title} {\enquote {\bibinfo {title}
  {Photochemically induced acousto-optics in gases},}\ }\href@noop {}
  {\bibfield  {journal} {\bibinfo  {journal} {Phys. Rev. Appl.}\ }\textbf
  {\bibinfo {volume} {22}},\ \bibinfo {pages} {024014} (\bibinfo {year}
  {2024})}\BibitemShut {NoStop}%
\bibitem [{\citenamefont {Michine}\ \emph {et~al.}(2024)\citenamefont
  {Michine}, \citenamefont {More},\ and\ \citenamefont {Yoneda}}]{MichinePF24}%
  \BibitemOpen
  \bibfield  {author} {\bibinfo {author} {\bibfnamefont {Yurina}\ \bibnamefont
  {Michine}}, \bibinfo {author} {\bibfnamefont {Richard~M.}\ \bibnamefont
  {More}}, \ and\ \bibinfo {author} {\bibfnamefont {Hitoki}\ \bibnamefont
  {Yoneda}},\ }\bibfield  {title} {\enquote {\bibinfo {title} {{Large-amplitude
  density waves produced in ozone-mixed gas by ultraviolet laser
  irradiation}},}\ }\href@noop {} {\bibfield  {journal} {\bibinfo  {journal}
  {Physics of Fluids}\ }\textbf {\bibinfo {volume} {36}},\ \bibinfo {pages}
  {041703} (\bibinfo {year} {2024})}\BibitemShut {NoStop}%
\bibitem [{\citenamefont {Ou}\ \emph {et~al.}(2025)\citenamefont {Ou},
  \citenamefont {Perez-Ramirez}, \citenamefont {Cao}, \citenamefont {Redshaw},
  \citenamefont {Wang}, \citenamefont {Dedeler}, \citenamefont {Lees},
  \citenamefont {Lancia}, \citenamefont {Oudin}, \citenamefont {Kur},
  \citenamefont {Mikhailova}, \citenamefont {Riconda}, \citenamefont {Michel},\
  and\ \citenamefont {Edwards}}]{OuSPIE25}%
  \BibitemOpen
  \bibfield  {author} {\bibinfo {author} {\bibfnamefont {Ke}~\bibnamefont
  {Ou}}, \bibinfo {author} {\bibfnamefont {Victor~M.}\ \bibnamefont
  {Perez-Ramirez}}, \bibinfo {author} {\bibfnamefont {Sida}\ \bibnamefont
  {Cao}}, \bibinfo {author} {\bibfnamefont {Caleb}\ \bibnamefont {Redshaw}},
  \bibinfo {author} {\bibfnamefont {Michelle~M.}\ \bibnamefont {Wang}},
  \bibinfo {author} {\bibfnamefont {Pelin}\ \bibnamefont {Dedeler}}, \bibinfo
  {author} {\bibfnamefont {Ben}\ \bibnamefont {Lees}}, \bibinfo {author}
  {\bibfnamefont {Livia}\ \bibnamefont {Lancia}}, \bibinfo {author}
  {\bibfnamefont {Albertine}\ \bibnamefont {Oudin}}, \bibinfo {author}
  {\bibfnamefont {Eugene}\ \bibnamefont {Kur}}, \bibinfo {author}
  {\bibfnamefont {Julia~M.}\ \bibnamefont {Mikhailova}}, \bibinfo {author}
  {\bibfnamefont {Caterina}\ \bibnamefont {Riconda}}, \bibinfo {author}
  {\bibfnamefont {Pierre}\ \bibnamefont {Michel}}, \ and\ \bibinfo {author}
  {\bibfnamefont {Matthew~R.}\ \bibnamefont {Edwards}},\ }\bibfield  {title}
  {\enquote {\bibinfo {title} {{Experimental creation of volume diffraction
  gratings in ozone using interfering ultraviolet lasers}},}\ }in\ \href@noop
  {} {\emph {\bibinfo {booktitle} {Optical Technologies for Inertial Fusion
  Energy}}},\ Vol.\ \bibinfo {volume} {13358},\ \bibinfo {editor} {edited by\
  \bibinfo {editor} {\bibfnamefont {Stavros~G.}\ \bibnamefont {Demos}}\ and\
  \bibinfo {editor} {\bibfnamefont {Carmen~S.}\ \bibnamefont {Menoni}}},\
  \bibinfo {organization} {International Society for Optics and Photonics}\
  (\bibinfo  {publisher} {SPIE},\ \bibinfo {year} {2025})\ p.\ \bibinfo {pages}
  {1335808}\BibitemShut {NoStop}%
\bibitem [{\citenamefont {Matteo}\ \emph {et~al.}()\citenamefont {Matteo},
  \citenamefont {Tochitsky}, \citenamefont {Mirov},\ and\ \citenamefont
  {Joshi}}]{MatteoCLEO25}%
  \BibitemOpen
  \bibfield  {author} {\bibinfo {author} {\bibfnamefont {D.}~\bibnamefont
  {Matteo}}, \bibinfo {author} {\bibfnamefont {S.}~\bibnamefont {Tochitsky}},
  \bibinfo {author} {\bibfnamefont {S.}~\bibnamefont {Mirov}}, \ and\ \bibinfo
  {author} {\bibfnamefont {C.}~\bibnamefont {Joshi}},\ }\bibfield  {title}
  {\enquote {\bibinfo {title} {Demonstration of efficient laser-induced {Bragg}
  grating in highly vibrationally excited {CO}$_2$ gas},}\ }in\ \href@noop {}
  {\emph {\bibinfo {booktitle} {CLEO 2025 conference, Long Beach,
  CA}}}\BibitemShut {NoStop}%
\bibitem [{\citenamefont {Garoz}\ \emph {et~al.}(2012)\citenamefont {Garoz},
  \citenamefont {González-Arrabal}, \citenamefont {Juárez}, \citenamefont
  {Álvarez}, \citenamefont {Sanz}, \citenamefont {Perlado},\ and\
  \citenamefont {Rivera}}]{GarozNF13}%
  \BibitemOpen
  \bibfield  {author} {\bibinfo {author} {\bibfnamefont {D.}~\bibnamefont
  {Garoz}}, \bibinfo {author} {\bibfnamefont {R.}~\bibnamefont
  {González-Arrabal}}, \bibinfo {author} {\bibfnamefont {R.}~\bibnamefont
  {Juárez}}, \bibinfo {author} {\bibfnamefont {J.}~\bibnamefont {Álvarez}},
  \bibinfo {author} {\bibfnamefont {J.}~\bibnamefont {Sanz}}, \bibinfo {author}
  {\bibfnamefont {J.M.}\ \bibnamefont {Perlado}}, \ and\ \bibinfo {author}
  {\bibfnamefont {A.}~\bibnamefont {Rivera}},\ }\bibfield  {title} {\enquote
  {\bibinfo {title} {Silica final lens performance in laser fusion facilities:
  {HiPER} and {LIFE}},}\ }\href@noop {} {\bibfield  {journal} {\bibinfo
  {journal} {Nuclear Fusion}\ }\textbf {\bibinfo {volume} {53}},\ \bibinfo
  {pages} {013010} (\bibinfo {year} {2012})}\BibitemShut {NoStop}%
\bibitem [{\citenamefont {Moir}(2000)}]{MoirFED00}%
  \BibitemOpen
  \bibfield  {author} {\bibinfo {author} {\bibfnamefont {R.W}\ \bibnamefont
  {Moir}},\ }\bibfield  {title} {\enquote {\bibinfo {title} {Grazing incidence
  liquid metal mirrors ({GILMM}) for radiation hardened final optics for laser
  inertial fusion energy power plants},}\ }\href@noop {} {\bibfield  {journal}
  {\bibinfo  {journal} {Fusion Engineering and Design}\ }\textbf {\bibinfo
  {volume} {51-52}},\ \bibinfo {pages} {1121--1128} (\bibinfo {year}
  {2000})}\BibitemShut {NoStop}%
\bibitem [{\citenamefont {Pierce}(2019)}]{PierceBook}%
  \BibitemOpen
  \bibfield  {author} {\bibinfo {author} {\bibfnamefont {Allan~D.}\
  \bibnamefont {Pierce}},\ }\enquote {\bibinfo {title} {Effects of viscosity
  and other dissipative processes},}\ in\ \href@noop {} {\emph {\bibinfo
  {booktitle} {Acoustics: An Introduction to Its Physical Principles and
  Applications}}}\ (\bibinfo  {publisher} {Springer International Publishing},\
  \bibinfo {address} {Cham},\ \bibinfo {year} {2019})\ pp.\ \bibinfo {pages}
  {583--648}\BibitemShut {NoStop}%
\bibitem [{Note1()}]{Note1}%
  \BibitemOpen
  \bibinfo {note} {For $t\gg \tau $, $h(u)H(t-u)\approx h(u)$, and the integral
  in Eq. \protect \eqref {eq:p1full} takes the form of a convolution,
  $h(t)*\cos (\Omega _{ac}t)/\tau $. Using the convolution theorem, $h(t)*\cos
  (\Omega _{ac}t)/\tau =\protect \mathcal {F}^{-1}\{ \protect \mathcal
  {F}[h(t)/\tau ] \protect \mathcal {F}[\cos (\Omega _{ac}t)] \}/\protect \sqrt
  {2\pi }$, with $\protect \mathcal {F}$ the Fourier transform (unitary
  definition, with angular frequency), and $\protect \mathcal {F}[h(t)/\tau
  ]=e^{-\omega ^2\tau ^2/2}$, $\protect \mathcal {F}[\cos (\Omega
  _{ac}t)]=\protect \sqrt {\pi /2}[\delta (\omega -\Omega _{ac})+\delta (\omega
  +\Omega _{ac})]$, leads to Eq. \protect \eqref {eq:p1late}.}\BibitemShut
  {Stop}%
\bibitem [{Note2()}]{Note2}%
  \BibitemOpen
  \bibinfo {note} {Here one must use the lowest order of the asymptotic
  expansion of the error function, $\protect \text {erf}(x) \approx 1-\exp
  [-x^2]/(\protect \sqrt {\pi }x)$ when $x\rightarrow \infty $.}\BibitemShut
  {Stop}%
\bibitem [{\citenamefont {Yariv}\ and\ \citenamefont {Yeh}(2002)}]{YarivBook}%
  \BibitemOpen
  \bibfield  {author} {\bibinfo {author} {\bibfnamefont {A.}~\bibnamefont
  {Yariv}}\ and\ \bibinfo {author} {\bibfnamefont {P.}~\bibnamefont {Yeh}},\
  }\href@noop {} {\emph {\bibinfo {title} {Optical Waves in Crystals:
  Propagation and Control of Laser Radiation}}},\ Wiley Series in Pure and
  Applied Optics\ (\bibinfo  {publisher} {Wiley},\ \bibinfo {year}
  {2002})\BibitemShut {NoStop}%
\bibitem [{\citenamefont {Moharam}\ and\ \citenamefont
  {Young}(1978)}]{MoharamAO78}%
  \BibitemOpen
  \bibfield  {author} {\bibinfo {author} {\bibfnamefont {M.~G.}\ \bibnamefont
  {Moharam}}\ and\ \bibinfo {author} {\bibfnamefont {L.}~\bibnamefont
  {Young}},\ }\bibfield  {title} {\enquote {\bibinfo {title} {Criterion for
  {Bragg} and {Raman-Nath} diffraction regimes},}\ }\href@noop {} {\bibfield
  {journal} {\bibinfo  {journal} {Appl. Opt.}\ }\textbf {\bibinfo {volume}
  {17}},\ \bibinfo {pages} {1757--1759} (\bibinfo {year} {1978})}\BibitemShut
  {NoStop}%
\bibitem [{\citenamefont {Oudin}\ \emph {et~al.}(2025)\citenamefont {Oudin},
  \citenamefont {Ghosh}, \citenamefont {Riconda}, \citenamefont {Lancia},
  \citenamefont {Kur}, \citenamefont {Ou}, \citenamefont {Perez-Ramirez},
  \citenamefont {Lee}, \citenamefont {Edwards},\ and\ \citenamefont
  {Michel}}]{OudinPOP25}%
  \BibitemOpen
  \bibfield  {author} {\bibinfo {author} {\bibfnamefont {A.}~\bibnamefont
  {Oudin}}, \bibinfo {author} {\bibfnamefont {D.}~\bibnamefont {Ghosh}},
  \bibinfo {author} {\bibfnamefont {C.}~\bibnamefont {Riconda}}, \bibinfo
  {author} {\bibfnamefont {L.}~\bibnamefont {Lancia}}, \bibinfo {author}
  {\bibfnamefont {E.}~\bibnamefont {Kur}}, \bibinfo {author} {\bibfnamefont
  {K.}~\bibnamefont {Ou}}, \bibinfo {author} {\bibfnamefont {V.~M.}\
  \bibnamefont {Perez-Ramirez}}, \bibinfo {author} {\bibfnamefont
  {J.}~\bibnamefont {Lee}}, \bibinfo {author} {\bibfnamefont {M.~R.}\
  \bibnamefont {Edwards}}, \ and\ \bibinfo {author} {\bibfnamefont
  {P.}~\bibnamefont {Michel}},\ }\bibfield  {title} {\enquote {\bibinfo {title}
  {{PIAFS}: A {2D} nonlinear hydrodynamics code to model gaseous optics},}\
  }\href@noop {} {\bibfield  {journal} {\bibinfo  {journal} {Physics of
  Plasmas}\ }\textbf {\bibinfo {volume} {32}},\ \bibinfo {pages} {072714}
  (\bibinfo {year} {2025})}\BibitemShut {NoStop}%
\bibitem [{\citenamefont {Edwards}\ \emph {et~al.}(2022)\citenamefont
  {Edwards}, \citenamefont {Munirov}, \citenamefont {Singh}, \citenamefont
  {Fasano}, \citenamefont {Kur}, \citenamefont {Lemos}, \citenamefont
  {Mikhailova}, \citenamefont {Wurtele},\ and\ \citenamefont
  {Michel}}]{EdwardsPRL22}%
  \BibitemOpen
  \bibfield  {author} {\bibinfo {author} {\bibfnamefont {M.~R.}\ \bibnamefont
  {Edwards}}, \bibinfo {author} {\bibfnamefont {V.~R.}\ \bibnamefont
  {Munirov}}, \bibinfo {author} {\bibfnamefont {A.}~\bibnamefont {Singh}},
  \bibinfo {author} {\bibfnamefont {N.~M.}\ \bibnamefont {Fasano}}, \bibinfo
  {author} {\bibfnamefont {E.}~\bibnamefont {Kur}}, \bibinfo {author}
  {\bibfnamefont {N.}~\bibnamefont {Lemos}}, \bibinfo {author} {\bibfnamefont
  {J.~M.}\ \bibnamefont {Mikhailova}}, \bibinfo {author} {\bibfnamefont
  {J.~S.}\ \bibnamefont {Wurtele}}, \ and\ \bibinfo {author} {\bibfnamefont
  {P.}~\bibnamefont {Michel}},\ }\bibfield  {title} {\enquote {\bibinfo {title}
  {Holographic plasma lenses},}\ }\href@noop {} {\bibfield  {journal} {\bibinfo
   {journal} {Phys. Rev. Lett.}\ }\textbf {\bibinfo {volume} {128}},\ \bibinfo
  {pages} {065003} (\bibinfo {year} {2022})}\BibitemShut {NoStop}%
\end{thebibliography}
\end{document}